\documentclass[a4paper,fleqn,usenatbib]{mnras}

\usepackage{mathptmx}

\usepackage[T1]{fontenc}
\usepackage{ae,aecompl}

\usepackage{graphicx}	
\usepackage{amsmath}	
\usepackage{amssymb}	

\usepackage{hyperref}
\usepackage{siunitx}

\title[Short title, max. 45 characters]{MNRAS \LaTeXe\ template -- title goes here}
\title[LOTUS]{LOTUS:  A low cost, ultraviolet spectrograph} 

\author[I. A. Steele et al.]{I. A. Steele,$^1$ 
J. M. Marchant,$^1$
H. E. Jermak,$^1$
R. M. Barnsley,$^1$
S .D. Bates,$^{1}$
\newauthor
N.R. Clay,$^{1}$
A. Fitzsimmons,$^2$
E. Jehin,$^3$
G. Jones,$^4$
C. J. Mottram,$^{1}$
\newauthor
R. J. Smith,$^1$
C. Snodgrass,$^5$
M. de Val-Borro,$^6$
\\
$^{1}$Astrophysics Research Institute, Liverpool John Moores University, L3 5RF, UK\\
$^{2}$Astrophysics Research Centre, School of Mathematics and Physics, Queen's University Belfast, BT7 1NN, UK\\
$^{3}$Institut d'Astrophysique et de G\'{e}ophysique, Universit\'{e} de Li\`{e}ge, all\'{e}e du 6 Ao\^{u}t 17, B-4000 Li\`{e}ge, Belgium\\
$^{4}$Mullard Space Science Laboratory, Department of Space \&
Climate Physics, University College London,
RH5 6NT, UK\\
$^{5}$Planetary and Space Sciences, Department of Physical Sciences, The Open University, Milton Keynes, MK7 6AA, UK\\
$^{6}$Department of Astrophysical Sciences, Princeton University, NJ
08544, USA
}

\date{Accepted 2016 May 26. Received 2016 May 26; in original form
  2015 December 10}

\pubyear{2016}

\begin{document}
\label{firstpage}
\pagerange{\pageref{firstpage}--\pageref{lastpage}}
\maketitle

\begin{abstract}
We describe the design, construction and commissioning of LOTUS; a simple,
low-cost long-slit spectrograph for the Liverpool Telescope.  The
design is optimized for near-UV and visible wavelengths and
 uses all transmitting optics. It exploits the instrument focal
plane field curvature to partially correct axial chromatic
aberration.  A stepped slit provides narrow ($2.5 \times 95$ arcsec) and
wide ($5 \times 25$ arcsec) options that are optimized for spectral resolution and flux
calibration
respectively.  On sky testing shows
a wavelength range of 3200--6300 {\AA} with a peak system throughput
(including detector quantum efficiency) of 15 per cent
and wavelength dependant spectral resolution
of $R=225-430$.  By repeated observations of the symbiotic emission line
star AG Peg we demonstrate the wavelength stability of the system
is $<2$ {\AA} rms and is limited by the positioning of the object in the
slit.  The spectrograph is now in routine operation monitoring the
activity of 
comet 67P/Churyumov-Gerasimenko during its current post-perihelion apparition.
\end{abstract}




\section{Introduction}
\label{sec:intro}  

In this paper we describe
the design, construction and commissioning of a simple spectrograph 
(LOTUS - LOw-cosT Ultraviolet Spectrograph) that provides sensitivity down to the atmospheric
cutoff at $\sim 3200$ {\AA}.  The project was particularly
inspired by the opportunity to make simultaneous ground based observations with the
in-situ measurements provided by the 2015 rendezvous of the {\em
  Rosetta} spacecraft \citep{rosetta1, rosetta2} with comet 67P/Churyumov-Gerasimenko.

Cometary nuclei have been stored at cold temperatures in 
the outer solar system for most of their lifetime.  They become active and 
release volatile species as they approach the Sun.  Several of these
species result in strong spectroscopic bands at near-ultraviolet (UV)
wavelengths that are inaccessible to many astronomical spectrographs
which often have wavelength ranges that extend only down to $\sim 3800$ {\AA}. 
The LOTUS design therefore particularly targets
the NH(0,0) band at $\sim3340$ {\AA} and the CN band at $\sim 3880$ 
{\AA}. NH is a granddaughter product of  NH$_3$ ice, a molecule with
astrobiological significance \citep{astrobiology}.  
CN gives an excellent overview of the relative level of total activity of
comets and can be compared with the total continuum
brightness to look at dust-to-gas ratios\citep{activity}.  By providing
some sensitivity at longer wavelengths (up to 6300 {\AA}) LOTUS also
allows observations of the strong C$_2$ and C$_3$ bands and 
potentially weaker features associated with species such as 
CO$+$ and [O{\sc i}].

\section{Optical Design}
\label{sec:optical}

   \begin{figure}
   \includegraphics[width=8.5cm]{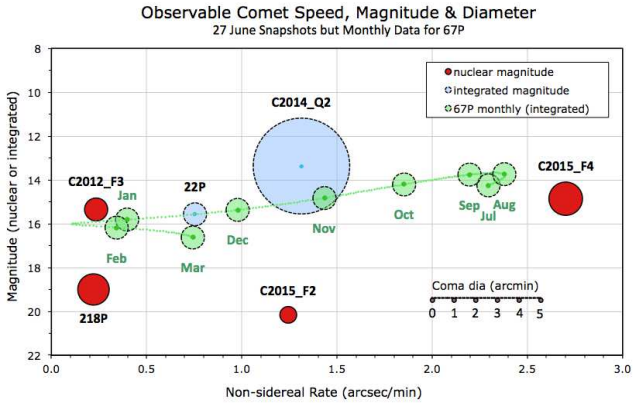}
   \caption 
   { \label{fig:coma} Predicted coma diameter, integrated visual magnitude
     and non-sidereal tracking rate for a typical selection of comets
     observable in late June 2015 from the JPL HORIZONS system.  
     For our principal target (67P)
     we present the evolution of these parameters in the
     post-perihelion season.  A slit length of 90 arcsec will be
     sufficient to fully sample the coma for the majority of targets.} 
   \end{figure} 

   \begin{figure}
   \includegraphics[angle=270,width=8.5cm]{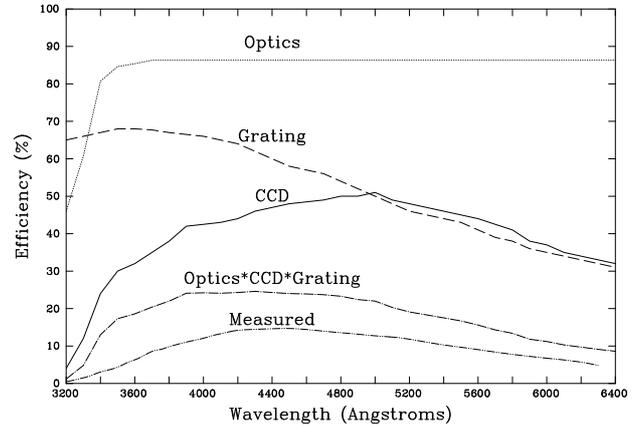}
   \caption 
   { \label{fig:ccd} CCD quantum efficiency (solid line), grating throughput (dashed
   line as a function of wavelength and optical throughput (dotted line).  The product 
   of the three is shown as a dot-dash line as is a prediction of the overall spectrograph
   throughput.  The throughput measured from observations of spectrophotometric standards
   is shown as a dot-dot-dash line (See section \ref{sec:throughput}). 
    } 
   \end{figure} 

\begin{table*}
\caption{Combined optical prescription of the Liverpool Telescope and
  LOTUS.  All dimensions in mm.}
\label{prescription}
\begin{tabular}{llllll}
\hline
Comment         & Type      &   Radius  & Thickness & Glass  &
Semidiameter \\
\hline
 Source                   & Standard  &      inf  &      inf  &      &    0.0000  \\
Primary                   &    Asphere&  -12000.0000   & 0.0000 &MIRROR & 1000.0000\\
                                & Standard    &    inf & -4315.3850  &    &   1000.0000\\
Secondary              &     Asphere & -4813.0000  &   0.0000& MIRROR  & 308.0000\\
                                & Standard     &   inf  &5600.0000 &      &   308.0000\\
Field Lens 250mm & Standard    &    inf    & 2.7100& F\_SILICA & 12.5000\\
Edmund 48-830    &Standard & -114.6200 &   13.0000   &    &    12.5000\\
                                &Standard    &    inf  & 198.0000  &     &    12.5000\\
Collimator 200mm &Standard    &    inf   &  2.8800 &F\_SILICA & 12.5000\\
Edmund 48-829    &Standard  & -91.6900   & 10.0000     &     & 12.5000\\
600l/mm (85-292) &Grating     &    inf   &  0.0000 &  &        12.5000\\
14 degree tilt      &Break         &  inf   &  0.0000    &    &  25.0000\\
                             &Standard   &     inf  &  10.0000  &  &       25.0000\\
Camera 125mm &    Standard  &  72.9500   &  5.0000& N-FK5  &   12.5000\\
Edmund 65-980  &  Standard  & -64.2100   &  2.5000& F2  &      12.5000\\
                 &Standard  &-181.9800 &  118.5000 &   &       12.5000\\
Detector     &    Standard &       inf    & 0.0000       &   & 18.0000\\
\hline
\end{tabular}
\end{table*}

   \begin{figure*}
   \begin{center}
   \begin{tabular}{c}
   \includegraphics[height=6.5cm]{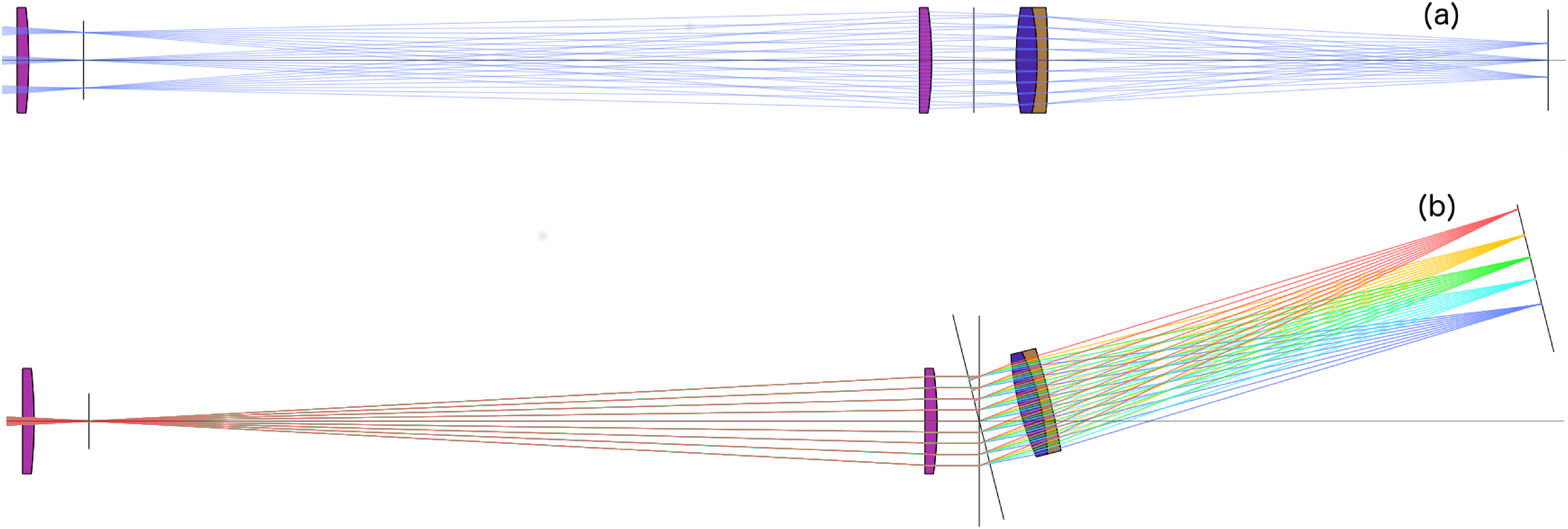}
   \end{tabular}
   \end{center}
   \caption 
   { \label{fig:layout} (a) LOTUS ray trace (side view) for on- and
     off- (1.2 arcmin) axis beams.  The field
     lens bends the incoming rays so the pupil is close to the
 succeeding collimator lens, grating and camera lens 
for both on- and off-axis rays and no
 vignetting occurs.  (b) LOTUS ray trace (top view).  In order of
 minimum to maximum deviation angle the rays are of wavelength 3200, 4000, 4700,
 5400 and 6200{\AA}.} 
   \end{figure*} 

LOTUS has been designed for use on the Liverpool Telescope (LT), 
although the
design should be easily adaptable to other telescopes of similar
focal ratio. 
LT \citep{steele2004} is a 2.0-m f/10 robotic (i.e. autonomous,
unmanned) telescope sited on the Spanish Canary Island of La Palma.
The telescope optical design is Ritchey-Chr\'{e}tien and provides a Cassegrain
focal station that can host up
to 9 instruments simultaneously.  The current instrument suite
comprises IO:O \citep{dummy} (an optical imaging camera), IO:I \citep{ioi1, ioi2} 
(an infrared imaging camera) , FRODOSpec \citep{frodo} (a medium
resolution optical spectrograph), SPRAT \citep{sprat} (a low resolution optical
spectrograph), RISE \citep{rise} (a fast readout imaging camera) and
RINGO3 \citep{ringo3} (a multi-band imaging polarimeter).  Swapping
between instruments is accomplished by motion of a retractable and
rotating fold mirror.  The instrument change time is $<30$ seconds.

The optical design of LOTUS is based on the requirement to obtain 
a wavelength range from 3200 {\AA} to at least 6000 {\AA} 
and a spectral resolution $R\sim300$.  This corresponds to velocity
resolution of $1000$ km/s.  In addition the design had to be 
sufficiently compact to fit easily on
the telescope, implying a length $<650$-mm, width $<300$-mm and
weight $< 25$-kg.

The slit length was determined by our wish to sample both the
nucleus and coma of our target comets.  In Fig. \ref{fig:coma} we
present the visual (similar to $V$-band) magnitude, non-sidereal 
rate and coma size for an example set of comets
observable in late June 2015 as well as C67P throughout its observing
season.  It can be seen that a slit length of 90 arcsec will allow
observations of the full coma of the majority of our potential
targets and we therefore adopted this as our design requirement.

The slit width was determined by the need to acquire the target
reliably in robotic (unattended) operation.  We use an automated iterative
imaging and world coordinate system (WCS) fitting procedure
on our main imaging camera (IO:O) to carry out acquisition.   We then
deploy the fold mirror to direct the beam to the spectrograph
once the target is in the correct focal plane position.
Experiments with the same mode of operation on
our optical SPRAT spectrograph \citep{sprat} showed that a projected
slit width of 2.5 arcsec was necessary to ensure reliable
acquisition.  

A 2.5 arcsec slit width on a 2.0-m telescope implies a collimated
beam width of 20-mm is necessary to achieve resolution
$R=300$ at a wavelength of 4500 {\AA} using a 600 line/mm
grating.  For the f/10 telescope beam, this implies a collimator
focal length of 200-mm.  The grating pitch 
was selected by comparing the efficiency curves of commercially 
available UV optimized gratings.  It was found that 600 line/mm has a sufficient groove
density to yield an acceptable resolution combined with reasonable
efficiency ($>30$ per cent) 
over the wavelength range of $3200-6400$ {\AA}  and
high efficiency
($>60$ per cent)  at the key wavelength range between 3200 and 4400 {\AA} 
(Figure \ref{fig:ccd}).

   \begin{figure*}
   \includegraphics[height=7cm]{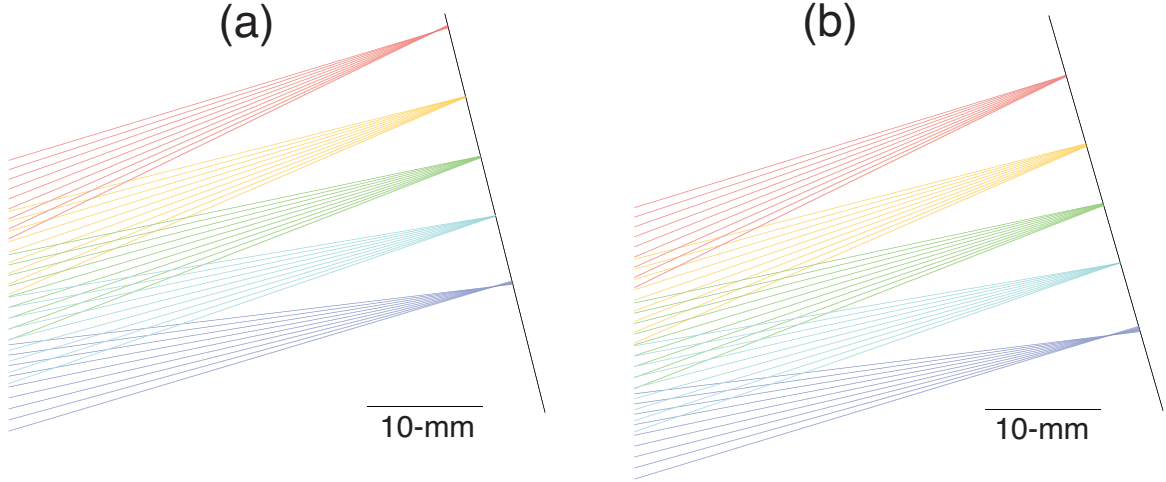}
   \caption 
   { \label{fig:zoom} (a) Detailed view of ray trace (wavelengths as
     per Figure \ref{fig:layout}) for the chosen system angle of 14$^\circ$ showing how by
     having the central wavelength displaced towards the upper section
   of the detector allows the field curvature to partially correct the
   axial chromatic aberration.  (b) As part (a) but for a system
   angle of 16.2$^\circ$ placing the spectrum symmetrically on the
   detector.  In this configuration the axial chromatic aberration is 
   unacceptably strong in the UV. } 
   \end{figure*} 

   \begin{figure*}
   \includegraphics[height=10cm]{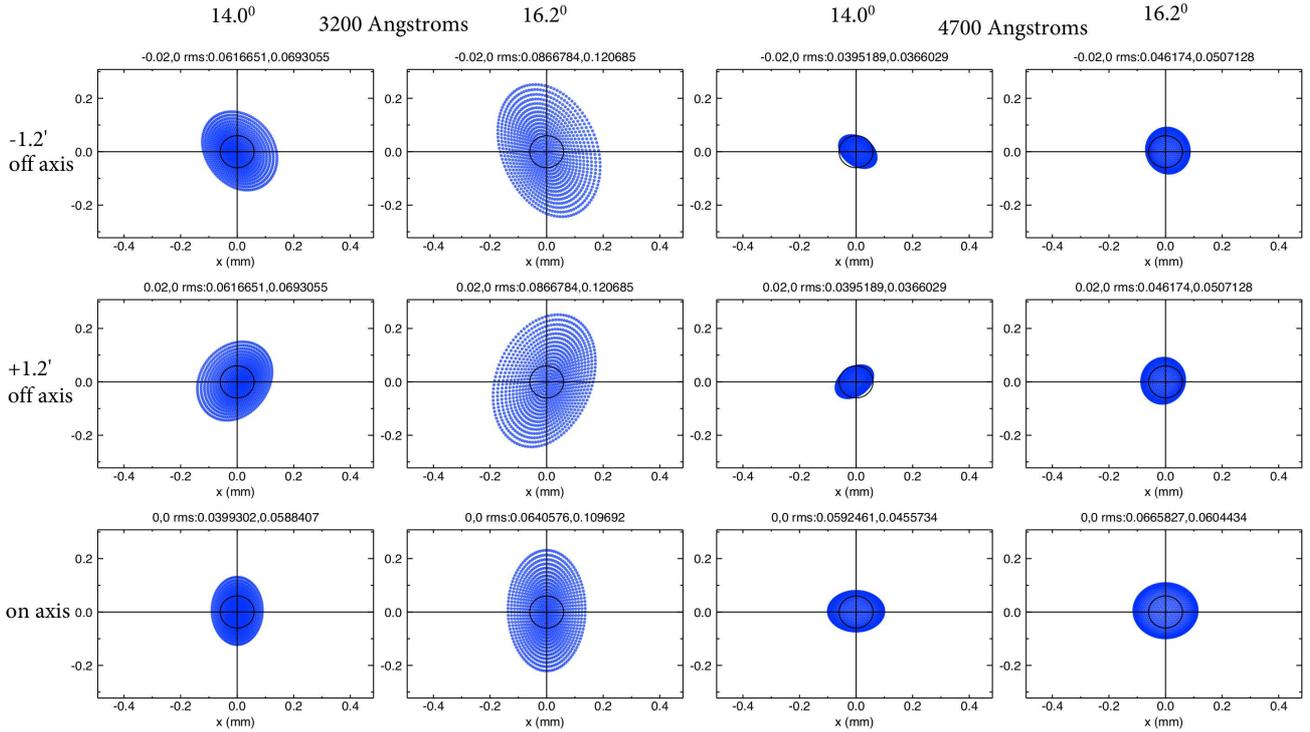}
   \caption 
   { \label{fig:spots} Spot diagrams for system angles of 14.0$^\circ$
     and 16.2$^\circ$.  The spots are generated assuming a point
     source (no atmospheric seeing) on axis and at $\pm1.2$ arcmin off
     axis.  Wavelengths of 3200 {\AA} and 4700 {\AA} are shown.  x and
     y axis are mm in the focal plane.  The black circle indicates a 2
     arcsec diameter.} 
   \end{figure*} 

   \begin{figure*}
   \includegraphics[height=7cm]{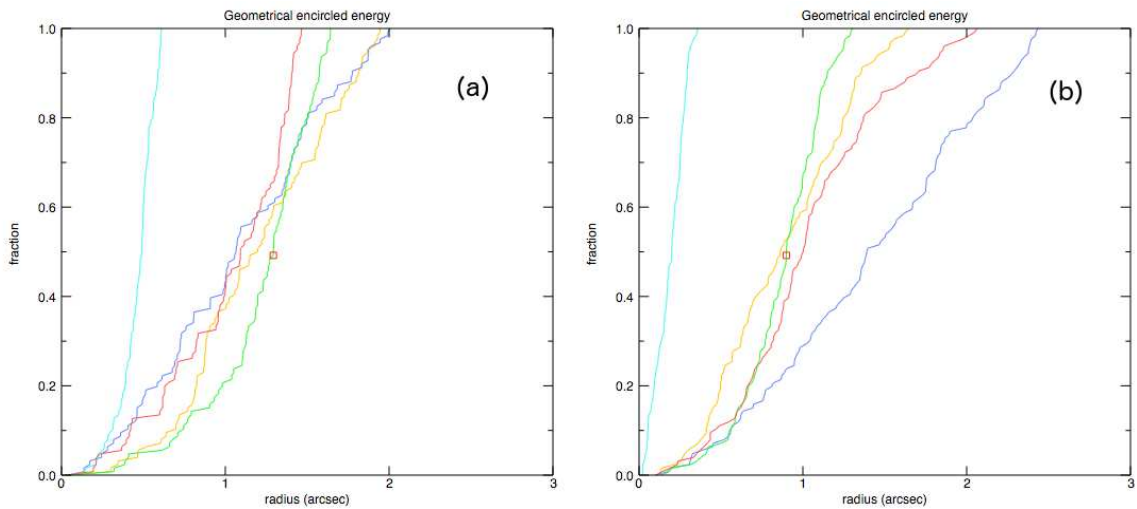}
   \caption 
   { \label{fig:gee} 
(a) Geometric Encircled Energy Plots on-axis for wavelengths (left to
right at GEE=1.0) of 4000, 6200, 4700, 5400 and 3200{\AA}.  (b)
Geometric Encircled Energy Plots 1.2 arcmin off-axis for wavelengths
(left to right at GEE=1.0) of 4000, 4700, 5400, 6200 and 3200 {\AA}.} 
   \end{figure*} 

   \begin{figure}
   \includegraphics[angle=270,width=8.5cm]{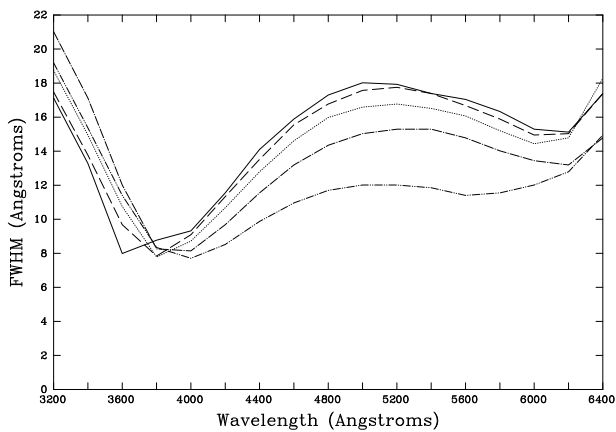}
   \caption 
   { \label{fig:res-calc} Calculated Full Width Half Maximum as a function of
   wavelength from our ray-tracing and assuming a seeing of 1.0 arcsec.  The calculation has been carried out at 
   wavelength increments of
   200{\AA} and on axis (solid line)
   and at angles of 20 arcsec (dashed), 40 arcsec (dotted), 60 arcsec (dot-dash) 
   and 80 arcsec (dot-dot-dash) off axis.
    } 
   \end{figure} 

   \begin{figure}
   \includegraphics[angle=270,width=8.5cm]{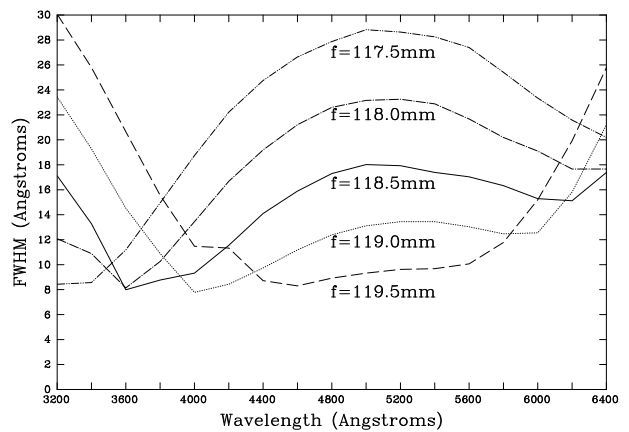}
   \caption 
   { \label{fig:focus-calc} Calculated Full Width Half Maximum as a function of
   focus distance (f) from the last surface of the camera lens to the detector 
   and assuming 1.0 arcsec seeing.  The calculation has been carried out at 
   the nominal focus (f=118.5mm) and values of $\pm0.5$- and $\pm1.0$-mm either 
   side.
    } 
   \end{figure} 

Since we were aiming for a low cost design, we limited our choice of
optical elements to those available off the shelf. The UV throughput 
requirement implies any optical elements used must avoid significant internal absorption or
air-glass surface reflections.  This ruled out the use of commercial
photographic lenses which have unknown prescriptions and coatings.  For single element
lenses, 
either fused silica or CaF$_2$ with an anti-reflection coating can be used.
However there is a much wider range of focal lengths and diameters available for fused silica 
and therefore this material was chosen.   
For achromatic lenses the choice was very
limited since the super-flint (SF) glasses typically used within them 
block light below $3700$ {\AA}.  It was possible however to identify a
small range of commercial  UV optimized achromats with relatively
large diameters (25 and 50-mm) using F2 glass which have reasonable transmission
down to $3200$ {\AA}.  

All lenses used in the spectrograph have a proprietary "UV-VIS" anti-reflection multi-coating 
which reduces reflection losses to $<1$ per-cent per surface between
3200 and 6400 {\AA}.
The total optical throughput (combining reflection and absorption losses) of the lenses
and the uncoated fused silica CCD window is also plotted in Figure \ref{fig:ccd}.

The optical design uses simple plano-convex lenses as 
a collimator and field lens and a UV achromat as a camera lens.  
A collimator focal
length f=200-mm and a camera focal length f=125-mm imply an image
magnification factor of 0.625 should be obtained on the detector.  This is slightly 
reduced to 0.587 by the 
250-mm plano-convex field lens which is placed just in front of the slit
to ensure that rays from the field edge pass through the
collimator and grating (Fig \ref{fig:layout}(a)).  
The system layout is shown in Figure \ref{fig:layout}(b).  The overall length of the optical path
from field lens to focal plane 
is $360$-mm.  Including the 75-mm distance from the telescope
mounting flange to the focal
plane and $100$-mm
 length of the detector body, the total instrument length is
 $535$-mm.  The design dispersion is $8$ $\mu$m/{\AA}, giving
a spectral length on the detector of $26$-mm.  The optical
prescription is presented in Table \ref{prescription}.

Ray tracing was carried out on-axis, at the field edge and at three intermediate
positions.  Fig.
 \ref{fig:zoom} highlights an important aspect of the design which
 partially corrects for axial (focus) chromatic aberration.  The
 natural spectrograph system angle is $16.2^\circ$.  This angle would
give a layout which places the spectrum symmetrically on either side of the
optical axis. However as shown in Fig. \ref{fig:zoom}(b) 
such a layout suffers from high axial chromatic aberration in the UV.  
The usual approach in this case is to correct the linear element of the 
axial chromatic aberration by tilting the detector with respect to the grating.  
However we found a better correction could be made by tilting both the grating and the detector 
to give a system angle of $14.0^\circ$. In this configuration  
(Fig. \ref{fig:zoom}(a)) we keep 
the grating and detector parallel but shift the centre of the spectrum off the 
optical axis.  The (usually undesirable) second order curvature of the
instrument focal plane symmetrical about the optical axis can then be used to achieve 
a better correction than a grating-only tilt.  For both system angles
the maximum negative (with respect to the focal plane) 
defocus occurs at 3200 {\AA}, however it is reduced from -2.7-mm 
at $16.2^\circ$ to -1.5-mm at $14.0^\circ$.    The maximum positive defocus is
similarly improved from +1.6-mm (at 4700 {\AA}) to +0.9-mm (at 5400 {\AA}).

In our design the detector is placed off 
centre with respect to the optical axis in the dispersion direction only.  This means
the offset does not significantly increase the effect of the field curvature in the spatial direction.
This effect is in any case lower since the  
maximum extent of the spectral image from the optical axis is only 3.7-mm in the spatial
direction, compared with 17-mm in the spectral direction.  
The lack of effect on the spatial imaging properties of the offset is also demonstrated 
in the spot diagrams for the two system angles presented in 
Fig. \ref{fig:spots}.  In our design
the 80 per cent geometric encircled energy diameter is $<3.2$ arcsec at all wavelengths on-axis,
and $<4.0$ arcsec off-axis (Fig. \ref{fig:gee}).  
Assuming a 2d Gaussian profile this corresponds to spatial full width half maxima (FWHM) of 
$<2.1$ and $<2.6$ arcsec respectively, well matched to the slit width of 2.5 arcsec. 

In Figure \ref{fig:res-calc} we plot the calculated spectral FHWM as a function
of wavelength at various field positions and assuming the input image
to the spectrograph has 1.0 arcsec seeing (FWHM).  The predicted 
FWHM varies between 21 {\AA} at 3200 {\AA} and FWHM=8 {\AA} at 4000 {\AA},
corresponding to a variation in spectral resolution from $R=150$ to $R=490$.  
Figure \ref{fig:focus-calc} shows
how the spectral FWHM varies as function of camera lens focus distance.  This 
variation is principally the effect of residual axial chromatic aberration and
demonstrates how our optimal focus ($f=118.5$-mm) is designed to give the best
compromise in terms of absolute FWHM (rather than resolution $R$) over the entire 
wavelength range.

\section{Detector}
\label{sec:detector}
Nearly all astronomical spectrographs use CCD detectors that are
thinned and back illuminated for best quantum efficiency (QE). 
Depending on the anti-reflection coating employed,
such detectors can have high QE over wide
wavelength ranges.  However such detectors are expensive and were
beyond the project budget.  We therefore employed a Starlight-Xpress
Trius-SX35 camera designed for
the amateur astronomy market. This is based on the use of a
Truesense KAI11002M interline CCD \citep{kodak}.  This detector has
$4032\times2688$  pixels of dimensions $9\times9$ $\mu$m.  
This yields a large image area of $36.3 \times 24.2$-mm.  
The CCD uses a microlens array to concentrate the light on the
active pixel areas.  
The microlens array and its associated glass cover plate have
a multi-layer Kodak proprietary anti-reflection coating.  The quoted QE
includes these coatings and is plotted in Figure \ref{fig:ccd}.
The detector has poor QE in the visible; however its UV sensitivity
(25 per cent QE at 3400 {\AA}) exceeds 
that of typical thinned, back illuminated detectors 
(15 per cent QE at 3400 {\AA}). It is therefore appropriate in this
application.   By using $4 \times 4$ binning we derive a 
spatial scale of $0.6$ arcsec/binned-pixel.  This matches
well the predicted image quality of 1.5 arcsec.  The potential
maximum slit length on the detector is $6.5$ arcmin, however 
vignetting by the field lens limits this to $2.4$ arcmin. 
In the wavelength direction the design gives a dispersion of 4.7{\AA} per
binned pixel.  This corresponds to 3.2 pixels per resolution element
at the nominal design $R\sim300$ and implying mild oversampling of the spectrum.

\section{Mechanical Design}
\label{sec:mech}

In order to aid rapid development at low-cost, the lenses and grating
were all housed in off-the-shelf T-mount aluminium optical tubes.  One tube
contains the field lens, slit, collimator lens and grating and the second
the camera lens.  Each T-mount tube had within it a finely threaded 
adjustment mechanism to allow collimator and camera focus.  The collimator
tube was mounted to the aluminium side plate of the spectrograph
body and the camera lens screwed to the front of the detector housing
via a thread adapter.  The collimator lens tube is
able to rotate within a locking mounting in order to align the slit
with the detector axis.   Similarly the grating is able to be locked and rotated
independently within the collimator tube to align it.  
While lacking any precision mechanism,
it was found that by making a series of small pseudo-random changes to these rotations 
during setup it was possible to align both the spatial 
and spectral directions with the edge of the detector to within
one binned pixel.  This corresponds to an accuracy of $<0.1^\circ$.
Finite element analysis
of the design predicted flexure of $<20 \mu$m at the focal plane for
a change in gravity vector of 70$^\circ$.
The direction of flexure (perpendicular to the beam) means that this
flexure should not affect wavelength calibration.

   \begin{figure*}
   \includegraphics[height=7cm]{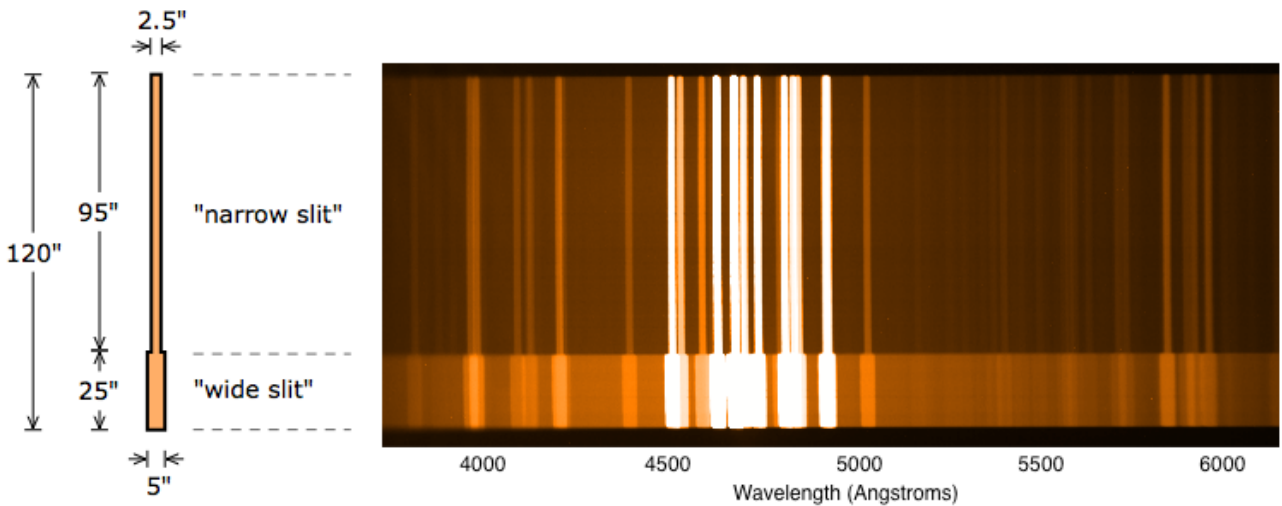}
   \caption 
   { \label{fig:slit} 
LOTUS slit design.  The slit consists of a narrow region for optimal
spectral resolution and a wide region for spectrophotometric
calibration.  Along side the slit is shown an exposure of a Xenon arc lamp.} 
   \end{figure*} 

To give the option of spectrophotometric calibration (which
requires a wide slit) we used spark erosion to cut a ``stepped''
slit in a thin aluminium disk. This gives a long (95 arcsec)
section of 2.5 arcsec projected width and a shorter
(25 arcsec) section with a projected width of 5.0 arcsec
(Fig. \ref{fig:slit}).
The total slit length is 2 arcmin, within the 2.4 arcmin limit
set by the system optics (Section \ref{sec:optical}).
By moving the target in the focal plane, the appropriate slit width can be
selected.

\section{Commissioning}
\label{sec:comm}
\subsection{Wavelength Calibration and Resolution}

   \begin{figure}
   \includegraphics[angle=270, width=8.5cm]{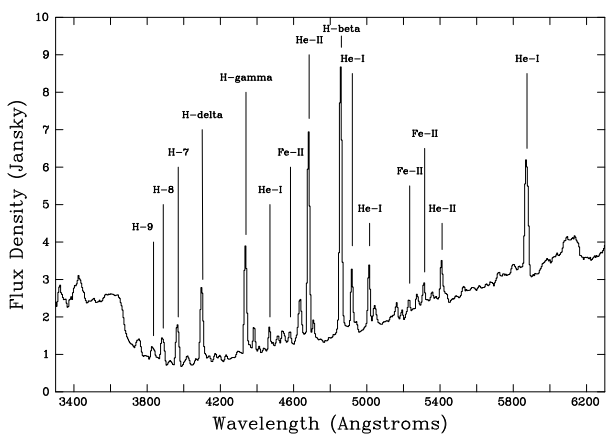}
   \caption 
   { \label{fig:agpeg} 
Spectrum of the symbiotic star AG Peg obtained on 2015 August 16.  The
spectrum has been wavelength calibrated using arcs taken prior to
installation.  The identified emission lines are used to check the
wavelength calibration (see text for details).  The spectrum has been
flux calibrated using an observation of the spectrophotometric
standard G191B2B and shows strong UV emission below the Balmer limit.}
   \end{figure} 

The spectrograph was installed on the telescope on 2015 June 29
following realuminzation of the telescope optics.
Since no calibration arc lamp unit is provided, immediately prior to
installation Xenon (Xe) and Neon (Ne) arc exposures were
taken on the bench.  The compact PenRay arc lamps were located $1-2$cm 
from the field lens, and provided even illumination (Fig. \ref{fig:slit}).  
The Xe and Ne exposures were coadded in software.  Out of the 23
  visible lines, 14 could be unambiguously identified against lines in
  the NIST atomic spectra database\footnote{http://www.nist.gov/}.  A second order polynomial
fit to those lines (which lay between 3370 and 6318 {\AA}) was 
made.
No systematic difference between the Xe and Ne lines was apparent. The measured linear dispersion was 
4.7 {\AA} per binned pixel as per expectation and the maximum deviation from linear
behaviour is 21 {\AA} at the extremes of the wavelength range.  The
second order fit residual has rms 1.8 {\AA}. This second order fit 
is used as the wavelength calibration for all subsequent on-sky data.

\begin{figure}
   \includegraphics[angle=0, width=7.5cm]{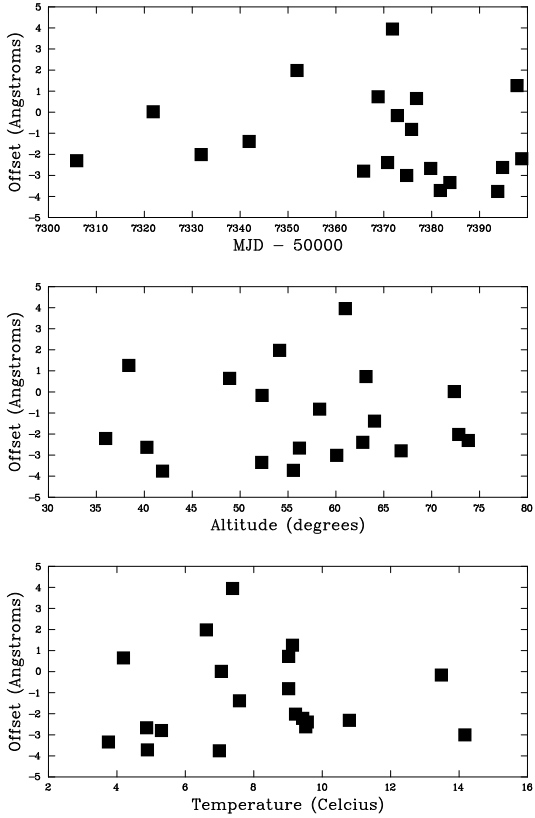}
   \caption 
   { \label{fig:cor} 
(top).  Offset in standard wavelength calibration derived from observations of
AG Peg as a function of date from 2015 Oct 26 to 2016 Jan 11.  No
trend is apparent. (middle) and (bottom) Wavelength offset as a
function of telescope altitude and tube temperature respectively for
data taken over the same date range.  Again no trends are apparent.}
   \end{figure}

In order to confirm the stability of the wavelength calibration, regular
observations were made of the symbiotic star AG Peg which
has a rich spectrum of optical and UV emission lines \citep{yoo2008}.
Nineteen observations were taken over the period 2015 Oct 26 to 2016 Jan 11
and the standard wavelength calibration applied (Fig. \ref{fig:agpeg}).  
Cross-correlation of the spectra showed
an rms variation in wavelength of 2.1 {\AA}, corresponding to $0.45$
binned CCD pixel.  This scatter is naturally
explained by the uncertainty of positioning of the target within
the 2.5 arcsec (i.e. 4 binned pixel) slit for an individual exposure.
We searched for correlations between the wavelength offset 
and various external parameters (Figure \ref{fig:cor}).
A Spearman rank test showed no evidence
of a correlation of offset with date ($\rho=-0.35,p=0.14$), telescope altitude
($\rho=0.10,p=0.68)$), or temperature ($\rho=0.12, p=0.63$).
We conclude that the dominant source of calibration uncertainty
appears to be the position of the object within the slit. 

\begin{figure}
   \includegraphics[angle=270, width=8.5cm]{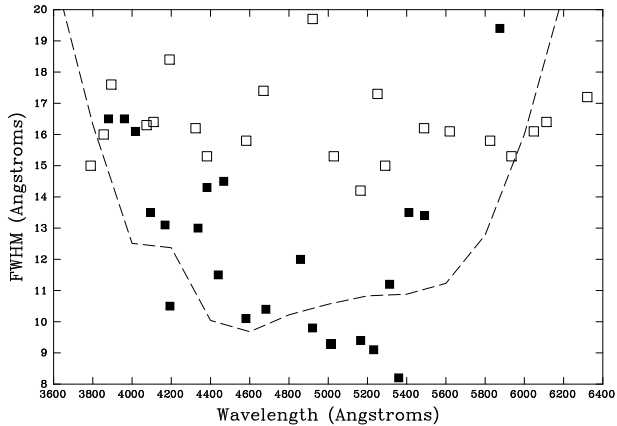}
   \caption 
   { \label{fig:res} 
Full Width Half Maximum (FWHM) of unblended arc lines (empty symbols) and stellar emission lines
(filled symbols) as a function of wavelength.  The FWHM for the arc lines is constant whereas that
for the stellar emission lines is smallest at the central region of the spectrum.  The dashed line
is the prediction from ray-tracing with seeing 1.2 arcsec and focus $f=119.5$-mm and shows the instrument is
operating approximately 1-mm from nominal focus.}
   \end{figure}

To assess the system resolution we used our bench arc spectrum from 2015 June 28 and an on-sky spectrum of 
AG Peg taken in good conditions (mean seeing 1.1 arcsec in $R$, 1.3 arcsec in $U$)
on the night of 2015 July 01.  We
measured the full width half maximum (FWHM) by Gaussian fitting 
of the brightest 22 unblended 
emission lines between 3700 and 6350 {\AA} in each spectrum.  
The results are shown in Figure \ref{fig:res}.  In the data derived from the
arc lines no trend is apparent. The mean FWHM of the arc lamp data is
16.3 {\AA}.  This appears to be a slight underestimate of the true FWHM
for the arc lines due to their flat topped profile.  Manual
measurement of the lines (to the nearest integer pixel) indicates a
mean FWHM of 4 pixels, corresponding to 18.8 {\AA}.

The on sky FWHM show a distinct dependance on wavelength,
with best FHWM achieved in the centre of the wavelength range.    
The mean FWHM over the whole wavelength range is 12.5 {\AA}.  At 4500 {\AA} the 
FWHM is 10.5 {\AA} corresponding to $R=430$ and considerably
better than that measured using the arc lines.  At 3800 {\AA} the FWHM is 17 {\AA}, 
similar to that of the arc lines and corresponding to $R=225$.  Also plotted on
Figure \ref{fig:res} is the predicted resolution as a function of wavelength
from our earlier ray tracing analysis (Section \ref{sec:optical})
for a camera focus offset of +1.0mm.  This makes it clear that the ``improved''
on-sky resolution at central wavelengths is likely to be due to 
a combination of the star under-filling the 2.5 arcsec
wide slit and an offset of the camera lens focus from the optimum position.  This focus
offset means that best image quality is obtained at the central wavelength regions and the worst
at the wavelength extremes.  This demonstrates a weakness of the current design.  
The combination of a wider (than the seeing disk) slit and lack of on sky focus adjustment means 
that a focus derived on the bench from arc lamps that fill the slit will not necessarily 
be optimal.  Nevertheless the achieved resolution performance is sufficient for the science 
goals.  The focus will be adjusted at a future date once the current observation
programme is completed.

\subsection{Throughput}
\label{sec:throughput}

The system throughput was tested by observing the spectrophotometric
standard stars GD191B2 \citep{oke90} and BD+33$^\circ$ 2642 \citep{stone77}.
The observed spectra were converted from measured ADU counts per wavelength bin
to photon counts per second per Angstrom by application of the appropriate scaling
factors for exposure time, CCD gain and dispersion.   The catalogue spectra were
converted to the same units taking into account scaling factors for
the variation of photon energy with wavelength, the telescope effective area, the variation
of atmospheric extinction with wavelength at the observed airmass \citep{king}, 
and the telescope optical throughput \citep{smith}.  The ratio of the observed to catalogue
spectra then gives the throughput.  The results of 
this analysis are presented in Figure \ref{fig:ccd}.

The peak throughput of the spectrograph (including the 
CCD camera) is 15 per cent at a wavelength of 4500 {\AA}, compared to the 24 per cent
prediction from combining the manufacturers data for the grating, CCD camera.  
The reason for this discrepancy is unclear
however we note that the manufacturer's figures are not measurements of the supplied items,
but merely ``representative'' and so there may be some sample variation.  To discover the
true reason would require removal of the spectrograph from the telescope and disassembly for 
bench testing of the individual components.  This would have consequent disruption to the current 
observing programme.  However we may investigate the issue further once the current programme is completed.

\begin{figure}
   \includegraphics[angle=270, width=8.5cm]{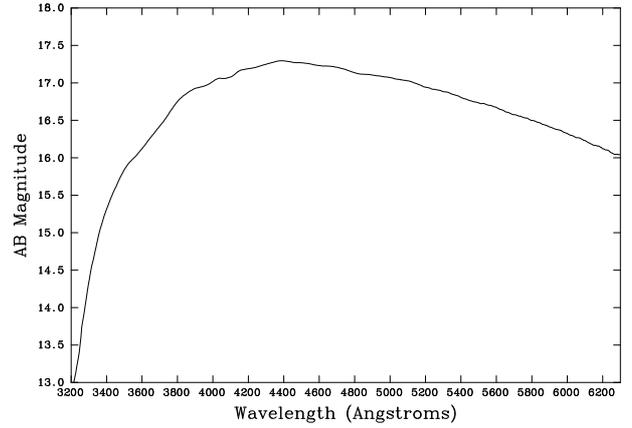}
   \caption 
   { \label{fig:abmag} 
AB Magnitude necessary to obtain one photon/sec/rebinned pixel with LOTUS.}
   \end{figure}

We also used our spectrophotometric standard observations to calculate the
sensitivity of the instrument expressed in terms of source AB magnitude that yields
one detected photon per second per wavelength bin (i.e. 4 binned pixels).  The results of this analysis
are presented in Figure \ref{fig:abmag} as an aid to planning observations with the
instrument.

\subsection{Ghost analysis}

No computational scattered light or ghost analysis was carried out during the system
design.  Empirical bench testing with a bright Neon arc lamp and
observations on sky with bright ($V\sim6$) stars shows a single curved ghost across the image.  
The ghost has a width of 10 binned pixels.  It has flux per image row of $<0.2$ per cent 
of the continuum spectrum.  We attribute the ghost to reflection of the zero order light 
by a curved surface in the spectrograph.  In normal use of the spectrograph 
it is not detectable.

\section{Observations}

   \begin{figure}
   \includegraphics[height=6.5cm]{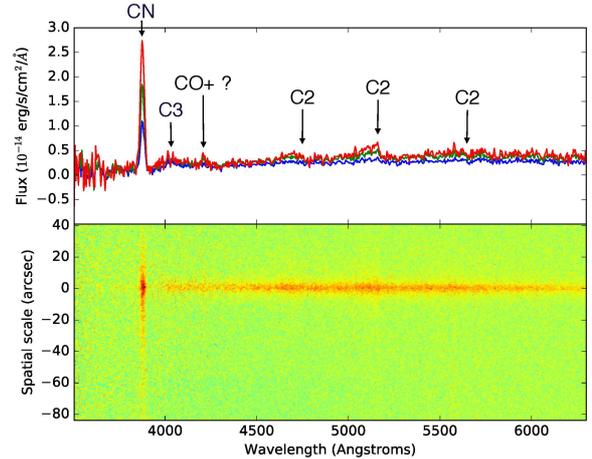}
   \caption 
   { 
1d and 2d spectra of comet 67P obtained on 2015 September 15 with
LOTUS.  The total exposure time is 900 s and an offset sky has been
subtracted. The 1d spectra are extracted at three different 
apertures
centred on the comet nucleus with diameters 8\farcs4 (blue line), 16\farcs8
(red line) and 33\farcs6 (green line). The comet is observed at a
heliocentric distance of 1.308 AU and the geocentric distance is 1.783 AU.} 
\label{fig:comet}
   \end{figure} 

As described in the introduction, LOTUS was developed
as a fast track instrument
to allow observations of comet 67/P over a range of heliocentric
distances in parallel with in-situ measurements from the {\em
  Rosetta} spacecraft.  A worldwide campaign of ground-based 
observations supports this mission, to provide large-scale context 
and a link to observations of other
comets\footnote{http://www.rosetta-campaign.net/}. 
Regular spectroscopic monitoring to measure gas production rates 
is a key element to this campaign, and one particularly suited to
robotic 
telescopes like the LT equipped with LOTUS.  This is especially 
the case during the months after the comet's perihelion 
passage in August 2015 when it was visible for only short periods 
pre-dawn each night.  We have therefore observed comet 67P repeatedly with 
LOTUS since September 2015.
We used an
ephemeris generated from the JPL Horizons system for non-sidereal tracking.
We also made regular observations of
flux standards (Section \ref{sec:throughput}) and a solar analogue star (HD29641).
A detailed analysis of our observations 
is in preparation which will combine data from other ground based
facilities and the in-situ spacecraft measurements.  
Here we simply present data from one epoch as a proof of concept.

In Figure \ref{fig:comet} we present an example spectrum 
obtained on 2015 September 15.  
The routine reduction pipeline 
is very similar to that of other long slit instruments.
Bias and dark frame
subtraction have been applied and the default wavelength calibration
added.   
Three spectra, each of 300-s exposure time
were obtained, plus a single matching offset sky exposure.  
Measuring a flat, featureless region of continuum between 4650 and 4720 {\AA} 
in a single spectrum extracted over a 10 pixel (6 arcsec)
aperture we found a signal-to-noise ratio (SNR) of 14.4 per binned pixel.  
The calculated SNR (based on photon counting statistics and using
the CCD read-noise of 23 electrons/pixel) is 13.8, in good agreement with the
measured value.

The three 300-s comet spectra obtained on the night 
were combined and extracted by summing the flux over an aperture along the 
slit and subtracting the background sky frame observed at the same
airmass 
and exposure time directly after the comet observations.
The spectra of the comet and the HD29641 solar analog were corrected 
for atmospheric extinction and flux calibrated using standard IRAF
techniques to obtain the spectra as flux versus wavelength. The
continuum in the comet spectra (due to 
reflection of solar radiation off the dust in the coma) was removed
using the spectrum of the solar analogue.

Apparent in both the 1d and 2d
spectra are strong CN emission at 3880 {\AA} as well as features
attributable to the C$_2$ ($\sim 4738,  5165,  5635$ {\AA}) bandheads.
The spatial extent of these
features compared with the continuum emission is also clear.
Also present (although weak) are C$_3$ ($\sim 4050$ {\AA}) and a
feature at 4200 {\AA} which could potentially be attributable to CO$+$.
We fitted a Gaussian function to measure the intensity of the CN line.
For an aperture with a diameter of 16\farcs8 a line flux of
\SI[separate-uncertainty=true]{1.85(11)e-13}{erg\per\s\per\cm\squared} was measured.
Based on comparing the flux from the comet in the IO:O acquisition images
with integrated spectra from the LOTUS instrument we concluded that there were no
significant flux losses when we use the 2.5 arcsec slit.

To convert the measured fluxes from the comet emission bands into production 
rates we used a simple Haser spherically symmetric model 
\citep{1957BSRSL..43..740H}. This model takes into account the release of 
parent molecules flowing outwards from the nucleus and photo-production and 
dissociation of daughter molecules with a characteristic scale-length. 
Despite of its simplicity the Haser model is widely used
to derive production rates of parent and daughter species,
and is useful to compare with the results from other observations.

Due to the faintness of the emission features in comet 67P we
only consider the CN emission in the analysis presented here.
We adopted the CN scale-lengths and fluorescence efficiencies from 
\citet{2010AJ....140..973S}. Using this model the column density for each 
of the extracted spectra is proportional to the measured flux of the emission 
band assuming that the comet is perfectly centred on the slit. 
A column density of \SI[separate-uncertainty=true]{4.98(55)e9}{\per\cm\squared} is derived.
The CN production rate was fitted simultaneously to the column densities 
obtained at three different apertures along the slit using a linear 
least-squares method. 
By fitting the column densities
at these positions in the coma up to a distance from the nucleus of
$2\times10^4$ km we find a CN production rate of \num[separate-uncertainty=true]{3.3(2)e24} molecules/s.
A comparison with similar measurements in 67P obtained with other
instruments will be presented in a forthcoming paper.
This method has proven to be less sensitive to model assumptions 
than using the line fluxes measured with small apertures.
Thus the derived value of the CN production
rate corresponds to an approximate average value over the molecules
contained in the slit apertures which were released on
time scales that spanned several hours assuming a typical outflow
radial velocity scaling of $0.85 \times r_\mathrm{h}^{-0.5}$ km/s
where $r_\mathrm{h}$ represents the heliocentric distance.

\section{Conclusions}

We have described the design and commissioning of a simple, long-slit
optical-UV spectrograph for the Liverpool Telescope. The design
uses off the shelf optics in an all transmitting configuration and
an ``amateur'' market CCD camera.
The instrument meets its design specification with a wavelength 
range of 3200--6300 {\AA} and a wavelength dependent spectral 
resolution $R=225-430$.
The wavelength calibration is stable to $2${\AA} rms and the 
combined instrument and detector peak throughput is 15 per
cent.  The instrument was designed, built and commissioned in 6 months at a
total cost of $<$\pounds10,000.  The design is sufficiently simple
that is should be straightforward to adopt to other telescopes at similar low
cost.  LOTUS is now in routine use monitoring the CN production rate in comet 67P.  

\section*{Acknowledgements}

The Liverpool Telescope is operated on the island of La Palma by
Liverpool John 
Moores University in the Spanish Observatorio del Roque de los
Muchachos of the 
Instituto de Astrof\'{i}sica de Canarias with financial support from the 
UK Science and Technology Facilities Council.  Ray tracing was carried out using
{\sc LensForge} by Ripplon Software Inc.









\bsp	
\label{lastpage}
\end{document}